\documentclass[aps,prl, reprint, showpacs, floatfix]{revtex4-1}
\usepackage{amsmath}
\usepackage{amssymb}
\usepackage{graphicx}
\usepackage{xcolor}
\usepackage{multirow}
\usepackage{mathtools}

\newcommand{\figref}[1]{Fig.~\ref{#1}}
\newcommand{\eqnref}[1]{Eq.~\eqref{#1}}

\newcommand {\doe}[2]{\ensuremath{\frac{\partial {#1}}{\partial {#2}}}}

\newcommand{\ket}[1]{\ensuremath{|{#1}\rangle}}

\newcommand{\vect}[1]{\ensuremath{\mathbf{#1}}}
\newcommand{\beq}{\begin{equation}}
\newcommand{\eeq}{\end{equation}}

\newcommand{\nfour}{\ensuremath{^{14}}N }

\begin{document}

\title{Measurement of the Berry Phase in a Single Solid-State Spin Qubit}
\author{Kai Zhang}
\author{N. M. Nusran}
\author{B. R. Slezak}
\author{M. V. Gurudev Dutt}
\email[Correspondence to:]{gdutt@pitt.edu}
\affiliation{Department of Physics and Astronomy, 3941 O'Hara Street, Pittsburgh PA 15260}	

\date{\today}

\begin{abstract}
We present measurements of the Berry Phase in a single solid-state spin qubit associated with the nitrogen-vacancy center in diamond. Our results demonstrate the remarkable degree of coherent control achievable in the presence of a highly complex solid-state environment. We manipulate the spin qubit geometrically by careful application of microwave radiation that creates an effective rotating magnetic field, and observe the resulting phase via spin-echo interferometry. We find good agreement with Berry's predictions within experimental errors. We also investigated the role of the environment on the geometric phase, and observed that unlike other solid-state qubit systems, the dephasing was primarily dominated by fast fluctuations in the control field amplitude.
\end{abstract}
\maketitle

Geometric and topological phases are exciting and unique features of quantum physics, with historical origins going back several decades\cite{Berry84}. These phases have been found to play a role in a diversity of physical phenomena, for instance in analogues of the Aharanov-Bohm effect\cite{Aharanov59}, the Pancharatnam phase for rotation of light polarization in twisted optical fibers\cite{Chiao86,Chiao88}, in close connection to gauge theories of quantum fields~\cite{Wilczek84}, and condensed matter physics e.g. in the anomalous Hall effect and fractional statistics of the quantum Hall effect\cite{Zhang05,Xiao10}. 

In the field of quantum information science, holonomic quantum computation\cite{Zanardi99, Pachos00,Sjoqvist08} was proposed to take advantage of the geometric phase which is impervious to certain types of errors. Geometric quantum logic gates were realized first with nuclear spins in liquid solutions of molecules using NMR techniques\cite{Jones00}. Geometric phase has also been observed with single superconducting qubits\cite{Leek07} and trapped ions\cite{Leibfried03}. Isolated electron and nuclear spins from donors or quantum dots in a semiconductor are one of the original archetypes for solid-state quantum information processing\cite{Kane98, Loss98}, but so far there has been no measurement of the Berry phase with such a single solid-state spin qubit. 

Nitrogen-vacancy (NV) centers in diamond are among the most promising candidates for quantum information applications, highly sensitive nanoscale quantum sensors, biological markers, and as single photon emitters in quantum communication\cite{Doherty13}. Our measurement of the Berry phase occurs in the complex solid-state non-Markovian environment experienced by spins in diamond. A critical question that we probe experimentally in our work is the effect of this environment on geometric phase and the corresponding decay of this phase. Previous theoretical work in adiabatic and non-adiabatic quantum control settings\cite{Chiara03,Carollo03,Zhu02}, and other solid-state qubits such as quantum dots\cite{SanJose06,Whitney04} and superconducting qubits\cite{Cucchietti10,Lombardo14} shows that geometric phase can be modified significantly by the environment. Our results are important to understand the effect of noise on geometric quantum logic gates within small quantum registers consisting of the NV center and proximal nuclear spins\cite{Dutt07,derSar12, Zu14}, similar to demonstrations in trapped ions\cite{Leibfried03} and superconducting qubits~\cite{Abdumalikov13}. Proposals to measure geometric phase in rotating diamond crystal\cite{Maclaurin12}, and for application in gyroscopes\cite{Ajoy12,Ledbetter12} also motivate our experiments.

\begin{figure}[hbt]
\scalebox{1}[1]{\includegraphics[width=3.4in]{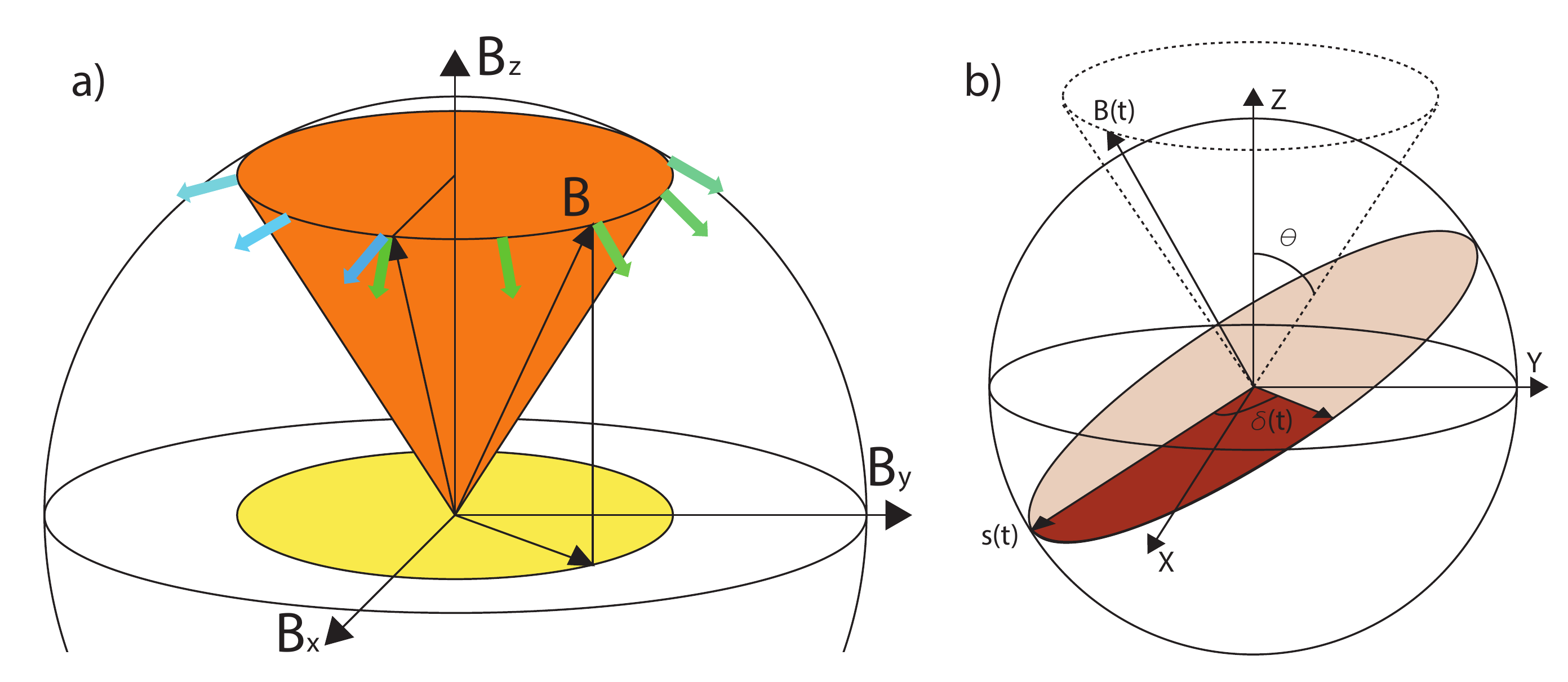}}
\caption{\small{
(a) Schematic illustration of the geometric phase accumulated by the spin vector (blue/green arrows) as it is transported along a closed path by the magnetic field applied to the spin qubit. (b) Bloch sphere picture of the precessing spin vector $\vect{S}(t)$ at one particular instant of time during the path of the magnetic field. }}
\label{figExp}
\end{figure}

In quantum mechanics, the Hamiltonian for a spin-$1/2$ qubit interacting with an external magnetic field is given by $H = \hbar \gamma_s \vect{B}\cdot \vec{\boldsymbol{\sigma}}/2$, where $\vec{\boldsymbol{\sigma}} = (\sigma_x, \sigma_y, \sigma_z)$ are the Pauli operators, $\hbar$ is Planck's constant, $\gamma_s$ is the appropriate gyromagnetic ratio for the particle, and $\vect{B}$ is the magnetic field vector. In the Bloch sphere picture (\figref{figExp}(b)), the qubit state $\vect{S} = \hbar \vec{\boldsymbol{\sigma}}/2$ continually precesses about the vector $\vect{B}$ acquiring dynamical phase $\beta_d (t) = \gamma_s B t$ where  $B = \lvert \vect{B} \rvert$. When the direction of $\vect{B}$ is now changed adiabatically in a time, i.e. at a rate slower than $\gamma_s B$, the qubit additionally acquires Berry phase while remaining in the same superposition with respect to the quantization axis $\vect{B}$. When $\vect{B}$ completes the closed circular path C shown in \figref{figExp}(a) the geometric phase acquired by an eigenstate is $ \pm \Theta_C/2$ where $\Theta_C$ is the solid angle of the cone subtended by $C$ at the origin.\cite{Berry84} The $\pm$ sign refers to opposite phases acquired by the ground or excited state of the qubit, respectively, giving rise to a relative geometric phase $\beta_g = \Theta_C$. For the circular path shown in the figure, the solid angle is given by $\Theta_C = 2 \pi (1 - \cos \theta)$, depending only on the cone angle $\theta$.    

\begin{figure}[hbt]
\scalebox{1}[1]{\includegraphics[width=3.4in]{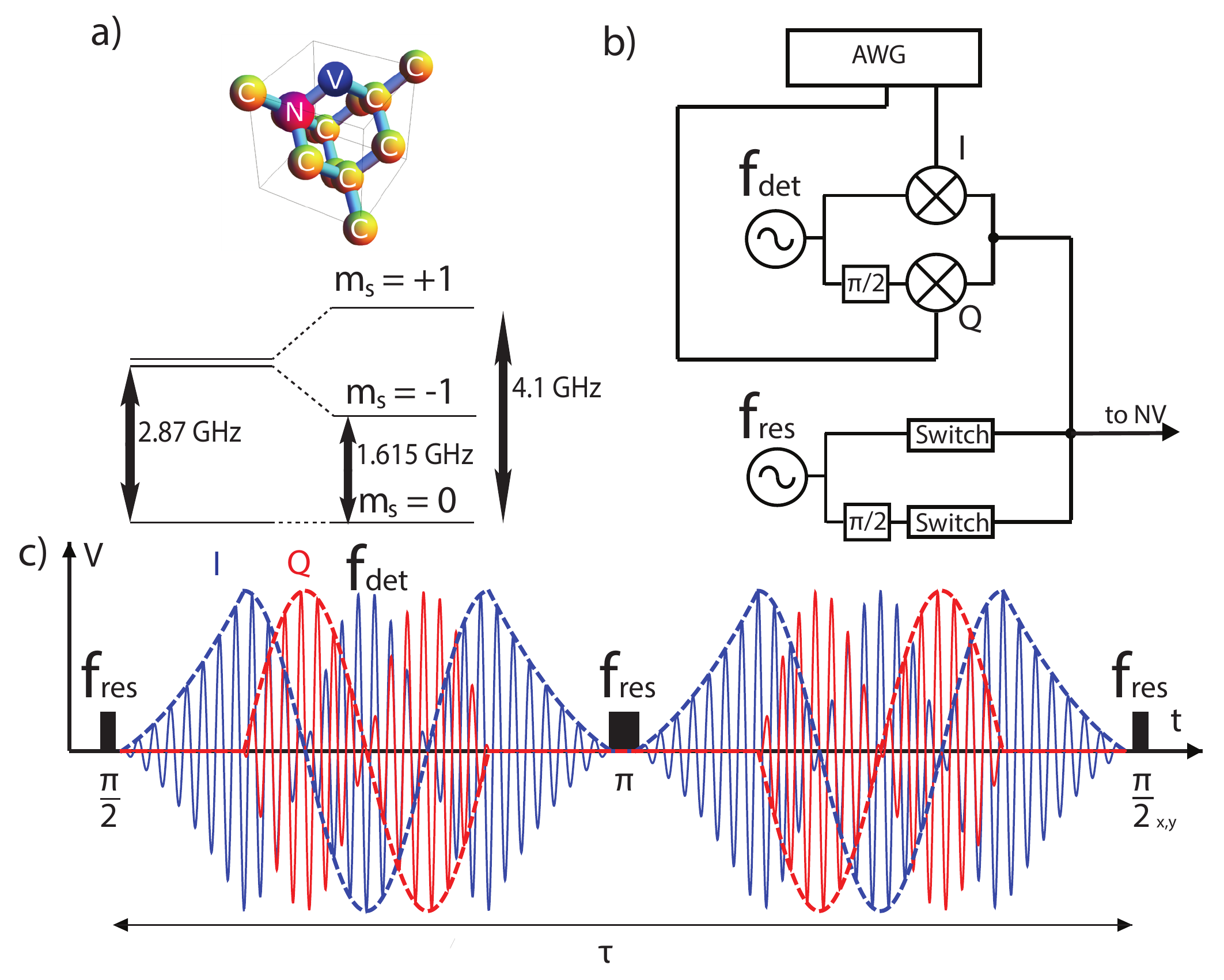}}
\caption{\small{
(a) (top) Lattice structure of the nitrogen-vacancy (NV) center in diamond showing substitutional nitrogen adjacent to a lattice vacancy.(bottom) Energy level diagram for the ground state of the NV center when a bias DC magnetic field near the excited state level anti crossing is applied. (b) Schematic circuit diagram showing the microwave synthesizers and arbitrary waveform generators (AWG) used for Berry phase experiments. (c) Pulse sequence (N=2) applied to the NV center combining both off-resonant drive at $f_{det}$ that causes geometric phase generation and resonant rotation pulses at $f_{res}$ used for a spin-echo interferometry sequence. The sequence time $\tau = 4\mu s$. The shape and timing of the ramp was chosen to best maintain adiabaticity~\cite{SM}.}}
\label{figSeq}
\end{figure}
The NV center (\figref{figSeq}(a)) is a spin-1 system in the ground state, quantized along the C$_{3v}$ symmetry axis between the nitrogen and vacancy sites, with the $\ket{m_s = 0}$ and $\ket{m_s = \pm 1}$ levels split by 2.87~GHz at zero magnetic field. The spin state can be initialized by optical pumping with 532 nm laser excitation, and the spin polarization can be detected by measuring the spin-dependent fluorescence signal. We use a single NV center in a type-IIa bulk diamond sample, and apply a static magnetic field $B_0$ oriented along the NV centers $z$-axis, allowing us to form a pseudo-spin $\sigma = 1/2$ qubit system with the $\ket{m_s = 0} \leftrightarrow \ket{m_s = -1}$ spin states. The magnetic field is chosen to bias the system near the excited-state level anti-crossing, resulting in complete polarization of the associated \nfour nuclear spin of the NV center\cite{Jacques09} and realizing a nearly ideal two-level system for our experiments. Microwave pulses are applied to the NV center using an impedance-matched microstrip line coupled to a thin copper wire on the diamond surface, allowing us to attain a $\pi/2$ rotation in $\sim 20$~ns. 

In the rotating frame of the microwave drive, and under the rotating wave approximation (RWA), the Hamiltonian for the NV center can be written as,
\beq
H = \hbar \Delta \sigma_z+ \hbar \Omega (\sigma_x \cos \Phi + \sigma_y \sin \Phi) 
\label{eq:hrwa}
\eeq
Here $\Delta$ is the detuning between the microwave and the transition frequency, $\Omega$ is the Rabi frequency of the on-resonance drive field and $\Phi$ is an adjustable control phase of the microwave. In the rotating frame, we can immediately identify the effective magnetic field as given by $\vect{B} = (\Omega \cos \Phi, \Omega \sin \Phi, \Delta)$. In our experiments, we typically keep $\Delta$ fixed and trace circular paths with different radii $\Omega$. The cone angle in our experiment would therefore by given by $\cos \theta = \Delta / (\Omega^2 + \Delta^2)^{1/2}$.

The corresponding adiabatic circuit for the magnetic field can be achieved by fast amplitude and phase modulation of the magnetic field\cite{Jones00,Leek07}, as shown in \figref{figSeq}(b). We measure Berry's phase using a spin-echo interference experiment. We apply two types of microwaves to the system as shown in \figref{figSeq}(c): one at a frequency $f_{det}$ that carries out the adiabatic circuit for $\vect{B}$, and another tuned to the resonance transition frequency $f_{res}$ that is used for the spin-echo interference, and subsequent state tomography to extract the spin vector. The spin-echo sequence initializes the qubit into an eigen-state of $\sigma_x$ with a resonant $\pi/2$ pulse.  The effective field $\vect{B}$ created by the off-resonant drive traces out a contour $C$ in \figref{figExp}(a), and the direction is set by the phase modulation $\Phi$. The relative quantum phase between the states $\ket{0}$ and $\ket{1}$ during the adiabatic circuit is given by $\phi_{\pm} = (\beta_d \mp \beta_g) $ where $\pm$ denotes counter-clockwise (clockwise) direction. Due to the resonant $\pi$-pulse in the spin-echo sequence, the cumulative relative phase becomes $\phi = \phi_- - \phi_+ =  2 \beta_g$, and increasing the number of times ($N$) that we trace the contour gives us $ \beta_g = 	(N/2) \Theta_C$.  The resonant $\pi$-pulse also cancels any fluctuations in environmental or other fields that occur on timescales slower than the echo sequence, thereby extending the decoherence time from the Ramsey dephasing time $T_2^*$ to $T_2$.

At the end of the sequence, we can extract the values of the spin vector through quantum state tomography. The value of $\langle S_z \rangle$ can be obtained using either Rabi oscillations or rapid adiabatic passage experiment to calibrate our fluorescence levels~\cite{SM}. The $\langle S_x \rangle $ and $\langle S_y \rangle$ values can be obtained by applying a $\pi/2$ pulse around the different axes $x$ or $y$, serving as a tomography pulse. The experimental phase can then be extracted as $\phi = \arctan (\langle S_y \rangle/ \langle S_x \rangle)$. Even in the presence of decoherence, which is a minor effect in our experiments~\cite{SM}, the coherence in either $x$ or $y$ direction would be equally affected and thereby will not alter the geometric phase we measure from taking the ratio. The total pulse sequence time and the shape of the waveforms was chosen to both preserve adiabaticity and to allow for the local spin environment of our qubit to return to its original state~\cite{SM}. 

\begin{figure}[hbt]
\scalebox{1}[1]{\includegraphics[width=3.4in]{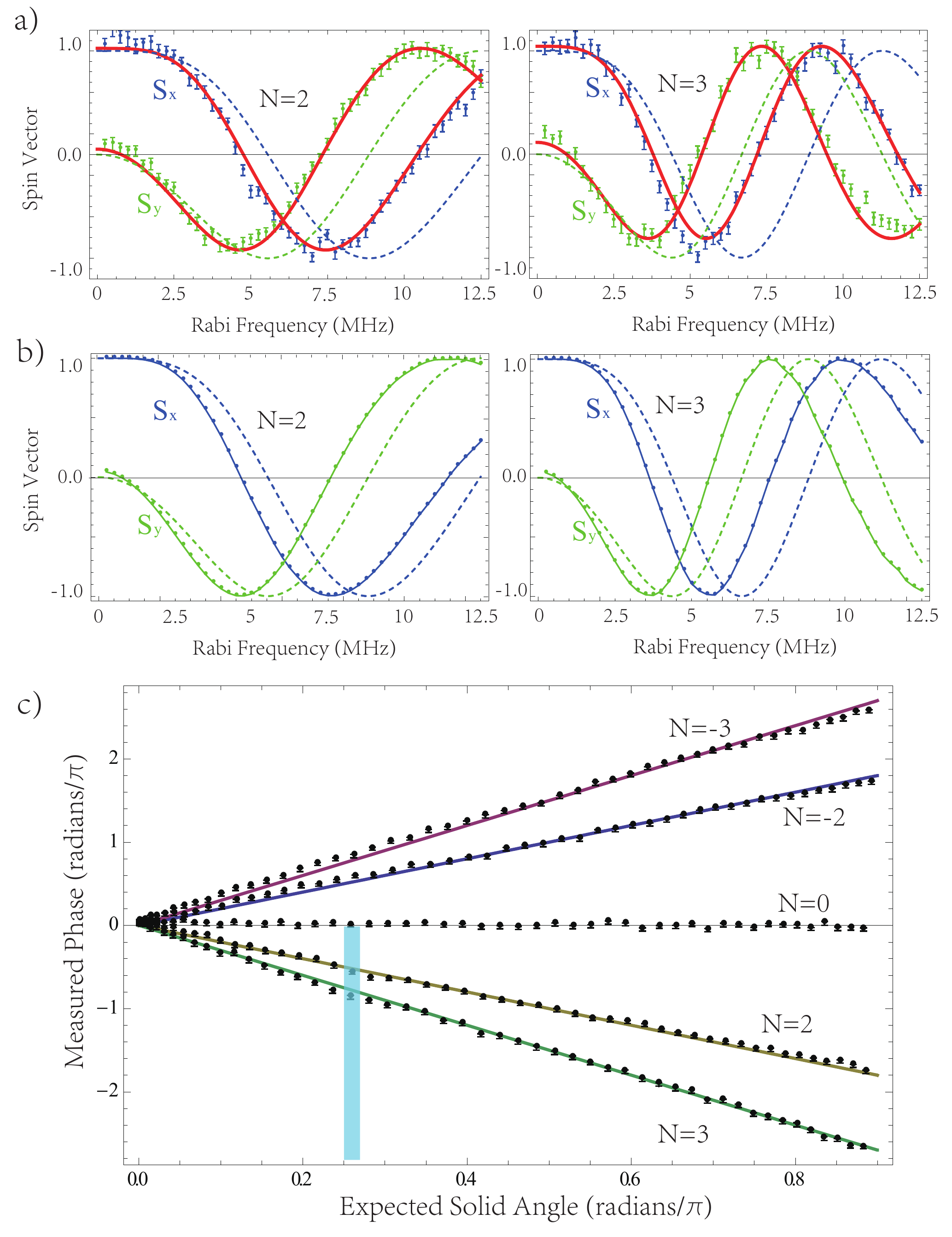}}
\caption{\small{
(a) Measured values of $\langle S_x \rangle$ and $\langle S_y \rangle$ from state tomography as a function of the Rabi frequency $\Omega$ for $\Delta = 10 \, \text{MHz}$. The fit functions (thick red lines) use a modified form of \eqnref{eq:prob} with $\Omega \rightarrow \Omega_{eff}$, while the blue (green) dashed lines are predicted directly for $\langle S_x \rangle$ ($\langle S_y \rangle$) by \eqnref{eq:prob} using the experimental parameters. (b) Numerical simulations using the density matrix formalism with blue (green) solid lines representing the predicted values of $\langle S_x \rangle$ ($\langle S_y \rangle$). The dashed blue (green) lines are the same as in part (a). The control field error values were measured independently and used as inputs for the simulation~\cite{SM}. (c) Plot of the measured quantum phase $\phi$ against the varying solid angle of the geometric circuit, for different values of $N$. Solid lines are straight lines with slope $\pm2, \pm3$, while data points were corrected for the effective $\Omega$ (see main text). Shaded region represents the solid angle at which the measurements of \figref{figDephasing} were carried out as a function of $N$. Each measurement point is obtained by repeating the experiment $ 2 \times 10^6$ times. }}
\label{figData}
\end{figure}
The data in \figref{figData}(a) shows the values of $\langle S_x \rangle $ and $\langle S_y \rangle$ measurements as the Rabi frequency $\Omega$ was varied while the detuning $\Delta$ remains fixed. The expected signal here from theory is, 
\begin{align}
\langle S_x \rangle &= A \cos \left(  2 \pi N ( 1 -  \frac{\Delta}{(\Omega^2 + \Delta^2)^{1/2}}) \right) \notag \\
\langle S_y \rangle &= - A \sin \left(  2 \pi N ( 1 -  \frac{\Delta}{(\Omega^2 + \Delta^2)^{1/2}}) \right)
\label{eq:prob}
\end{align}

The theoretical prediction shown in \figref{figData}(a) differs from the measured values in a systematic fashion that is at first puzzling. We identified the most significant deviation to be caused by single channel nonlinearities in our microwave circuitry, which distorts the experimental field path from the ideal geometric circuit shown in \figref{figExp} and changes the effective solid angle $\Theta_C \rightarrow \Theta_{C,eff}$. We verified that the deviations are not due to simple calibration errors in our Rabi frequency, and carried out a number of other control experiments and checks on our microwave parameters which were used as inputs for a full numerical simulation of the state evolution\cite{SM}. As shown in \figref{figData}(b), our simulations taking into account experimentally measured errors, and no other free parameters, show good agreement with the data and similar systematic deviation from the theory. To compare our data further to theory, we parametrized the solid angle $\Theta_{C,eff}$ by one free parameter $\Omega_{eff}$, since we found that the frequency $\Delta$ of the microwave drive field was very well characterized. We stress this effective Rabi frequency is not due to a calibration error in our microwave amplitude, but appears only in Berry phase experiments due to the distortion of the ideal Berry circuit. Our data is well fit to this simplified model as shown by the thick lines \figref{figData}(a).

When we correct the solid angle uniformly by this one fit parameter ($\Omega_{eff}$ at a fixed value of $\Delta$), we obtain nearly perfect agreement between theory and our measurements. In \figref{figData}(c) we plotted the values of the quantum phase versus the solid angle for different $N$. The $x$-axis was corrected using the fit parameter $\Omega_{eff}$. The data for $N = 0$ is also a confirmation that the spin-echo is highly successful in canceling any quantum phase accumulated in either half of the sequence. We have verified that the adiabaticity parameter $A = \dot{\Phi} \sin \theta/ \lvert \vect{B} \rvert \ll 0.1 $ in the regime of evolution times that were used in our experiments~\cite{SM}.   We should also note that a similar deviation from theory can be seen in another solid-state qubit system (Ref.\cite{Leek07}) although it was not explicitly identified as due to these imperfections in that work. 

\begin{figure}[hbt]
\scalebox{1}[1]{\includegraphics[width=3.4in]{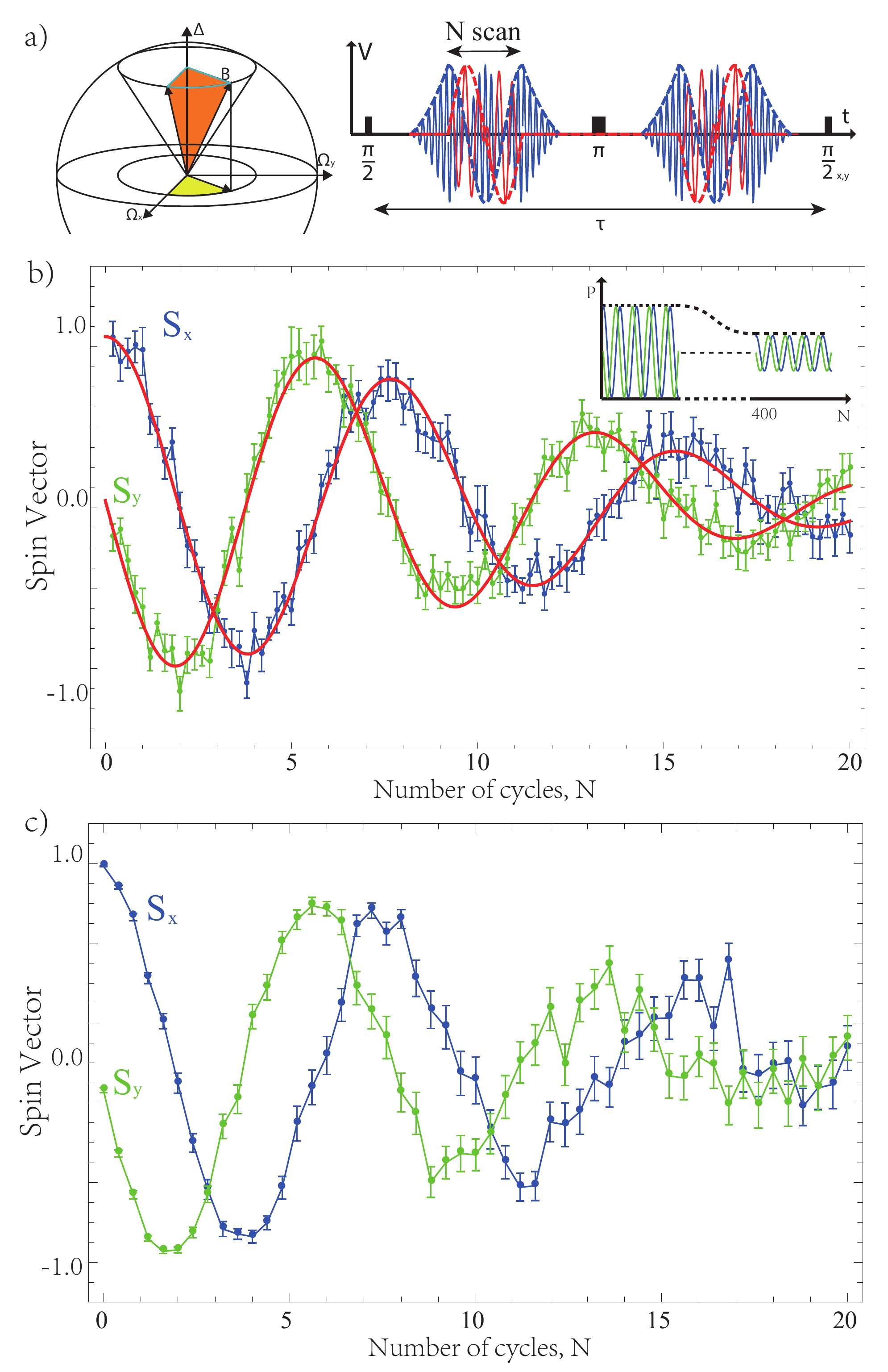}}
\caption{\small{
(a) (left) Path followed by effective field $\vect{B}$ for fractional circuits executed during $N$ scans.  (right) Microwave sequence for the $N$ scans. The sequence time $\tau = 20\mu s$. (b) Measured values of $\langle S_x \rangle$ and $\langle S_y \rangle$ from state tomography as a function of $N$ for fixed values of the solid angle $\Theta_C \sim 0.8$~radians, but with different detuning.  (c) Numerical simulations of $\langle S_x \rangle$ and $\langle S_y \rangle$ for the above sequence, assuming Gaussian amplitude fast fluctuations. See main text for explanation and interpretation.}}
\label{figDephasing}
\end{figure}
Our final set of measurements explores the question of how the geometric phase decays in the presence of the complex solid-state environment that interacts with the central NV spin.  Decay of the geometric phase is an important parameter to be measured for feasibility of geometric quantum information processing~\cite{Zanardi99,Pachos00} and for NV spin gyroscopes and mechanical rotation sensors~\cite{Maclaurin12,Ajoy12,Ledbetter12}. As shown in \figref{figData}(a), we were unable to detect any geometric dephasing over the solid angles obtainable by scanning the Rabi frequency $\Omega$. To increase the amount of geometric phase accumulated, we used the fact that the total quantum phase $\phi =  N \Theta_C$. Hence, we carried out an experiment with $N$ continuously varying while keeping the solid angle $\Theta_C$ fixed, using the pulse sequence in \figref{figDephasing}(a). 

Our experimental data in  \figref{figDephasing}(b) clearly showed decay as the amount of geometric phase is increased, which could be potentially explained as due to slow (adiabatic) fluctuations in either $\Delta$ or $\Omega$. Indeed this was the model used in Refs.~\cite{Carollo03, Chiara03, Leek07} to study geometric dephasing. We can estimate the strength of these fluctuations for $\Delta$ and $\Omega$ independently for our NV center from Ramsey and Rabi measurements~\cite{SM}. Assuming a gaussian model for the noise spectrum in $\Delta$, we obtain the following expression for the decay in the Berry phase as a function of the circuital number $N$:
\beq
\langle S_x \rangle = A\cos (N \Theta_C) \exp (- (N/N^*)^2) 
\eeq
where $N^* = \frac{\sqrt{2}}{\alpha \sigma_{\Delta}} = \frac{2 \pi T_2^*}{\alpha}$, $\sigma_{\Delta}$ is the standard deviation of the noise, $\alpha = \lvert \partial \Theta_C/ \partial \Delta \rvert = \frac{\Omega^2}{(\Omega^2 + \Delta^2)^{3/2}}$ . Given the numerical values in our experiments, we obtained $N^* \sim 400$, and generated the anticipated prediction (see inset to \figref{figDephasing}(b)). Thus, unlike the case of superconducting qubits, the slow noise in $\Delta$ from this environment is insufficient to explain the decay in our data~\cite{Leek07}. Similarly, our long-time Rabi oscillation data shows that the noise $\sigma_{\Omega}$ is even smaller and cannot explain this decay~\cite{SM}. 
The observed decay occurs on a timescale $\tau \ll T_2$ for our qubit, where the spin-echo sequence is expected to protect against fluctuations of $\Delta$. In fact, we can estimate the corresponding Berry phase fluctuations given the measured $T_2$ as well, and find that it is too small to explain the decay~\cite{SM}. We concluded that the decay is due to fluctuations in our microwave amplitude that cause the equatorial component of the field $\vect{B}$ to vary randomly from one half of the spin echo sequence to the other. We carried out numerical simulations of this experiment, assuming that the amplitude of the microwave field has a Gaussian distributed random component $\sim 10^{-3}$ within the echo sequence time. The results of these simulations with all other parameters taken from the experiment are shown in \figref{figDephasing}(c), and show good agreement with our experimental data. 

Our simulations and experiments suggest that we can attribute the decay in the experiments primarily to fluctuations of the dynamic phase caused by technical limitations of our AWG and microwave circuitry, which is not inherent to the NV spin qubit system. We present more evidence in \cite{SM} to support our model, such as measurement of dynamic phase, further measurements of geometric phase, and numerical studies. 

Our work reports on detailed measurements of the Berry phase in a single solid-state spin qubit, and on the effects of noise and control field imperfections on the Berry phase. Although the echo sequence cancels the average dynamic phase as well as slow dynamic phase fluctuations from the environment as shown in \cite{SM}, it cannot cancel the fluctuations from the microwave amplitude itself that occur on the timescale of the echo sequence. The contribution to the decay from radial fluctuations in the geometric path is significantly smaller, but could be important depending on the experimental parameters and geometric path followed. In general, we expect that any geometric quantum logic gate operation that relies on cancellation of the dynamic phase such as recently reported in Ref.~\cite{Zu14} will have to account for such control field fluctuations.



\begin{acknowledgments}
This work was supported by the DOE Office of Basic Energy Sciences (DE-SC 0006638) for development of geometric quantum control techniques, key equipment, materials and supplies, and effort. G.D. gratefully acknowledges support from the Alfred P. Sloan Foundation.
\end{acknowledgments}

\section*{Supplementary Material}

\setcounter{figure}{0}
\renewcommand{\thefigure}{S\arabic{figure}}
\section{NV Center ESR and Experimental Setup}

NV centers in our type-IIa single crystal diamond sample (sumitomo with [1 1 1] orientation) are located with our home-built confocal microscope. A high NA dry microscope objective (Olympus 0.95 NA) is used in this confocal setup for NV excitation and collection of fluorescence emission. Phonon-mediated fluorescent emission (650-750 nm) for the single NV center is detected under coherent optical excitation (LASERGLOW IIIB 532nm laser) using a single photon counting module(PerkinElmer SPCM-AQR-14-FC). Green excitation of the NV center polarizes the electron spin into $\ket{m_S = 0}$ sublevel of the $^3A_2$ ground state due to optical pumping. The rate of fluorescence signal counts varies for the $\ket{m_S = 0}$ and $\ket{m_S =\pm 1}$ states, which enables the optical detection of the electron spin. The photon counting takes place at the both ends of the optical excitation pulse(FIG.S1(b)). The counts at the beginning known as "signal"($S_i$) is highly dependant on the NV state while the counts at the end of optical excitation known as "reference" ($R_i$) is not. By taking the ratio of $S_i$ and $R_i$ as our fluorescence level (a.u.), we minimize the effect of laser fluctuations on our experiments. 
\begin{figure}[h!]
	\includegraphics[width=3.4in]{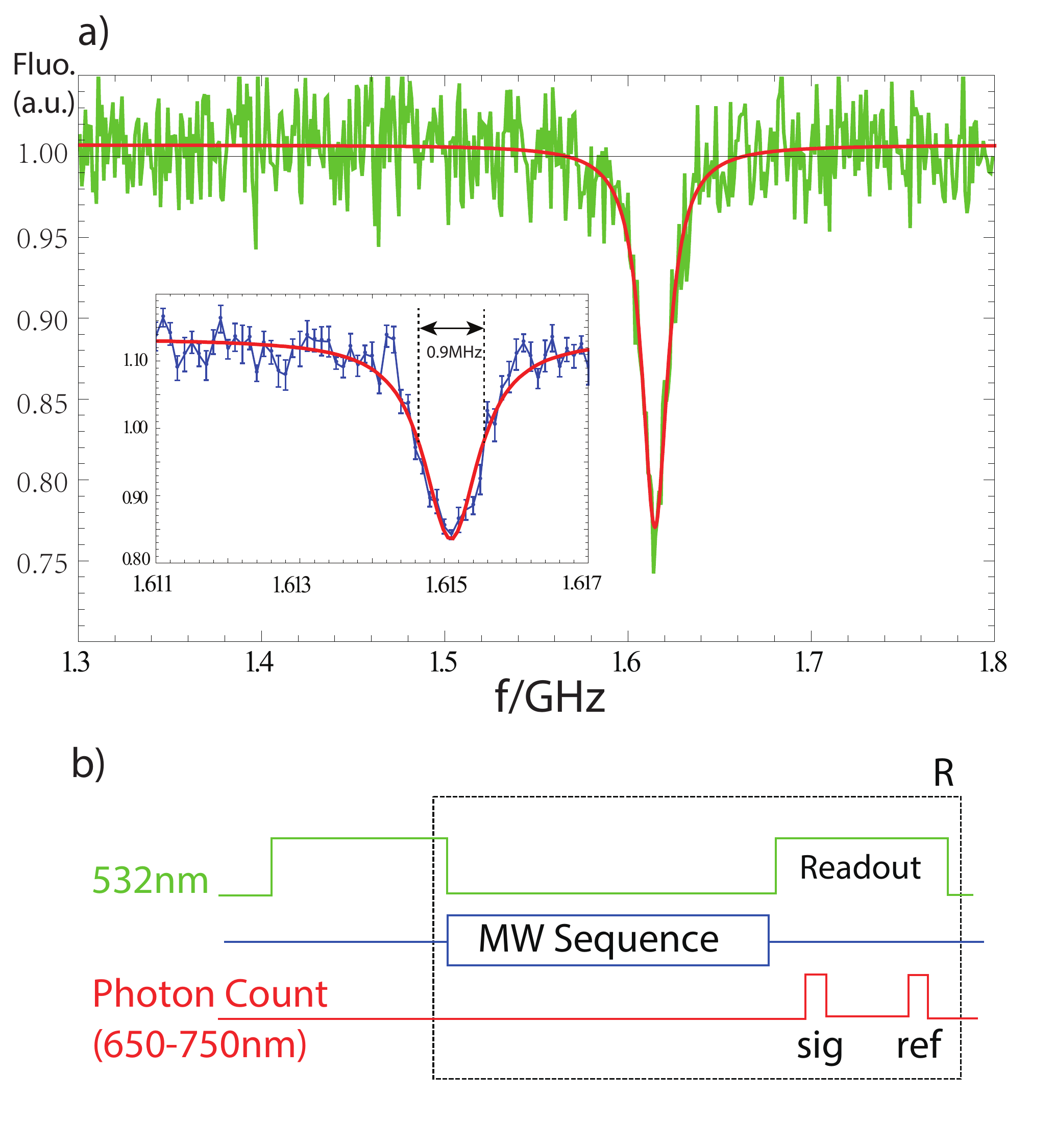}
	\caption{\textbf{a)} Optically Detected Magnetic Resonance of Single NV center. \textbf{Inset)} Pulsed ESR with scanning frequency. The linewidth of resonance dip is under 1MHz with 1 $\mu s$ pulse length. \textbf{b)} Schematic experimental sequence. The sequence within the dashed box is repeated 50000 times for one photon counting measurement.}
\end{figure}
A dc bias magnetic field $B_{0} \approx 450$~Gauss is applied along the NV axis by a permanent magnet. The transition frequency between $\ket{m_S = 0}$ and $\ket{m_S = -1}$ states is measured by Optically Detected Magnetic Resonance (ODMR) experiment and Pulsed ESR frequency scan experiment (FIG.S1 (a)).
Detuned microwave for Berry phase experiment is generated by Rohde$\&$Schwarz (SMIQ03B) signal generator with built-in IQ modulator. On resonance microwave is generated by PTS3200 synthesizer. Both microwaves are delivered via a 20-micron-diameter copper wire placed on the diamond sample. The input waveforms for IQ channels are generated by Tektronics AWG520 Arbitrary waveform generater (1G sample/s). The value of $\Delta$ is extremely well controlled in our experiments as we frequently track the resonance frequency of the NV center. Both synthesizers in our experiment are synchronized by an atomic clock and measured to drift less than $\sim 1 \times 10^{-3}\, \text{Hz}$, and this is tested by mixing the two microwaves from the two synthesizers under the same command frequency with a mixer and measuring the DC output of the mixer.

\section{Spin Echo Sequence: Larmor revivals}

The $^{13} C$ nuclear spin bath that has a natural abundance of $\approx 1.1 \% $ effectively produces a random field with frequency set by the nuclear gyromagnetic ratio $\gamma_n = 2 \pi (10.75)$~MHz/T and the dc bias field $B_0$. This random field causes collapses and revivals in the CP signals. For Berry phase experiment, it is required to operate on a revival point of spin echo sequence. In our Berry phase amplitude scan, the spin echo sequence time is set to the first revival. In N scan, the sequence time is set to the fifth revival.
\begin{figure}[h!]
	\scalebox{1}[1]{\includegraphics[width=3.4in]{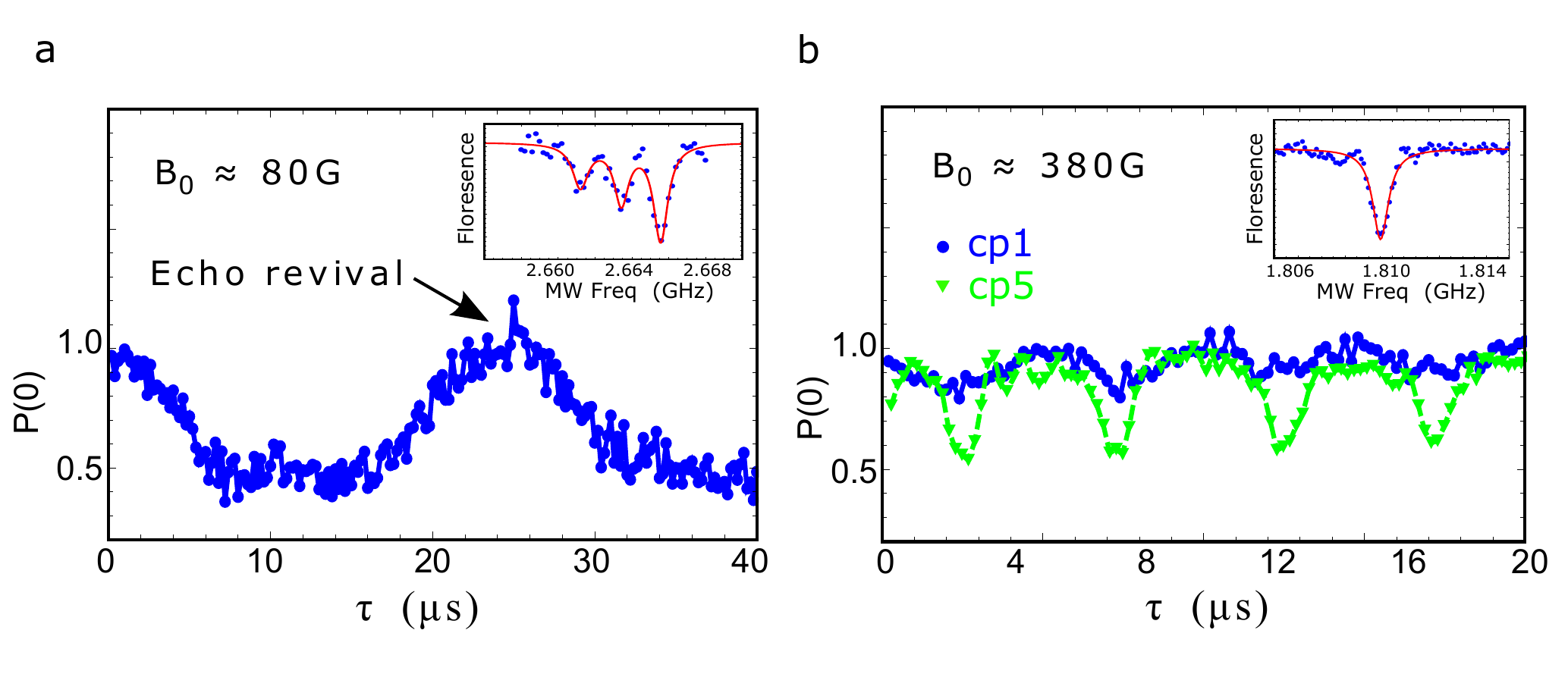}}
	\caption{\small{ Larmor revivals at \textbf{a},  $B_0 \approx 80$~G and  \textbf{b}, $B_0 \approx 380$~G. These revivals occur due to the effective random magnetic fields arising from Larmor precession of the $^{13} C$ nuclear bath that has a natural abundance of $\approx 1.1 \% $. The insets show the corresponding ODMR spectrum for those bias fields. Magnetic fields near excited state level anti-crossing causes dynamic nuclear polarization of $^{14}$N.}}
	\label{figRev}
\end{figure}

\section{Calibration Experiments}

\subsection{Fluorescence Level of \ket{0} and \ket{1} State}
In our experiments, the direct signal we were measuring is fluorescence counts. The fluorescence counts signal was mapped to probability in \ket{0} or \ket{1} state by calibration of fluorescence levels at \ket{0} and \ket{1} states with Rabi oscillation. However, imperfections in $\pi$ pulses might result in imperfect rotations and thereby cause errors in the fluorescence calibration.

\begin{figure}[h!]
	\includegraphics[width=3.4in]{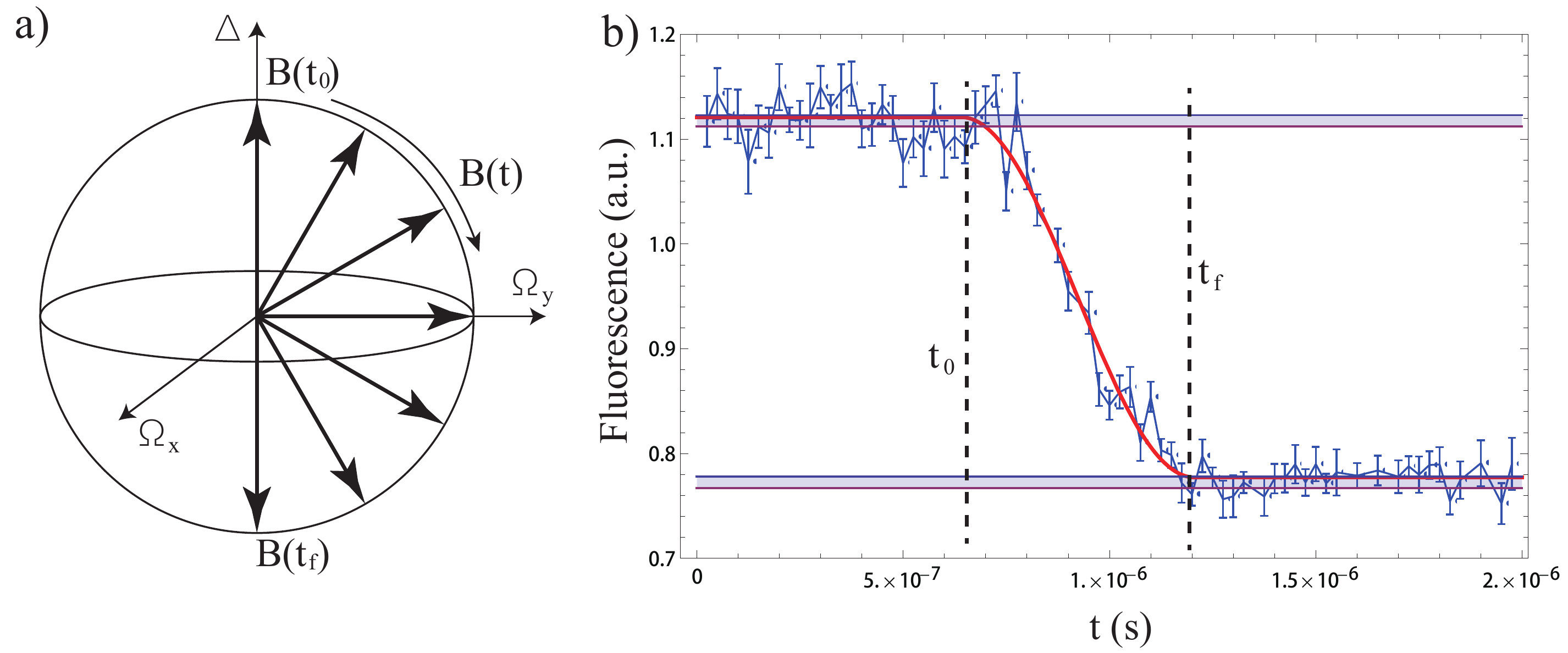}
	\caption{\textbf{a)} Schematic Time dependent effective B field. The effective B field slowly varies from $+z$ direction to $-z$ direction during the time from $t=t_0$ to $t=t_f$. \textbf{b)} Fluorescence measurement at time t.}
	\label{figPassage}
\end{figure}

Adiabatic passage is another way of calibrating fluorescence levels, and is used to check our Rabi oscillation calibration. By modulating both the amplitude and frequency of the driving microwave, an effective magnetic field adiabatically varying from $+z$ direction to $-z$ direction was created, as shown in \figref{figPassage}(a). The detuned microwave starts ramping up at $t=t_0$ while modulating its frequency, and then ramps back down to 0 with an opposite detuning at $t=t_f$. The magnitude of the effective B field is 12.5 MHz. Fluorescence level at different time t was measured and shown in \figref{figPassage}(b).

The data points before $t_{0}$ and after $t_{f}$ is fitted to fluorescence levels in \ket{0} and \ket{1} state resplectively. The fitted fluorescence levels fall in the confidence interval from our Rabi calibration, meaning our fluorescence level calibration is accurate.

\subsection{Microwave Power at Different Frequencies}

To test the frequency response of our microwave circuit, Rabi oscillation were measured under detuned microwave driving field with different detuning frequencies. Oscillations at 0MMHz, 10MHz and 20MHz detuning are shown as examples in FIG.S4 (a),(b) and (c). The fitted parameters, oscillation frequencies $\Omega(\Delta)$, amplitudes $amp(\Delta)$ and equilibrium position $bkg(\Delta)$ are plotted as functions of detuning frequency in FIG.S4 (d),(e) and (f).

The theoretical expectations are as following:
\begin{equation}
\Omega(\Delta)=\sqrt{\Omega(0)^2+\Delta^2}
\end{equation}
\begin{equation}
amp(\Delta)=amp(0)\frac{\Omega(0)^2}{\Omega(0)^2+\Delta^2}
\end{equation}
\begin{equation}
bkg(\Delta)=bkg(0)+amp(0)\frac{\Delta^2}{\Omega(0)^2+\Delta^2}
\end{equation}
The data demonstrates that our knowledge of MW parameters are accurate as a function of the microwave frequency.

\begin{figure}[h!]
	\includegraphics[width=3.4in]{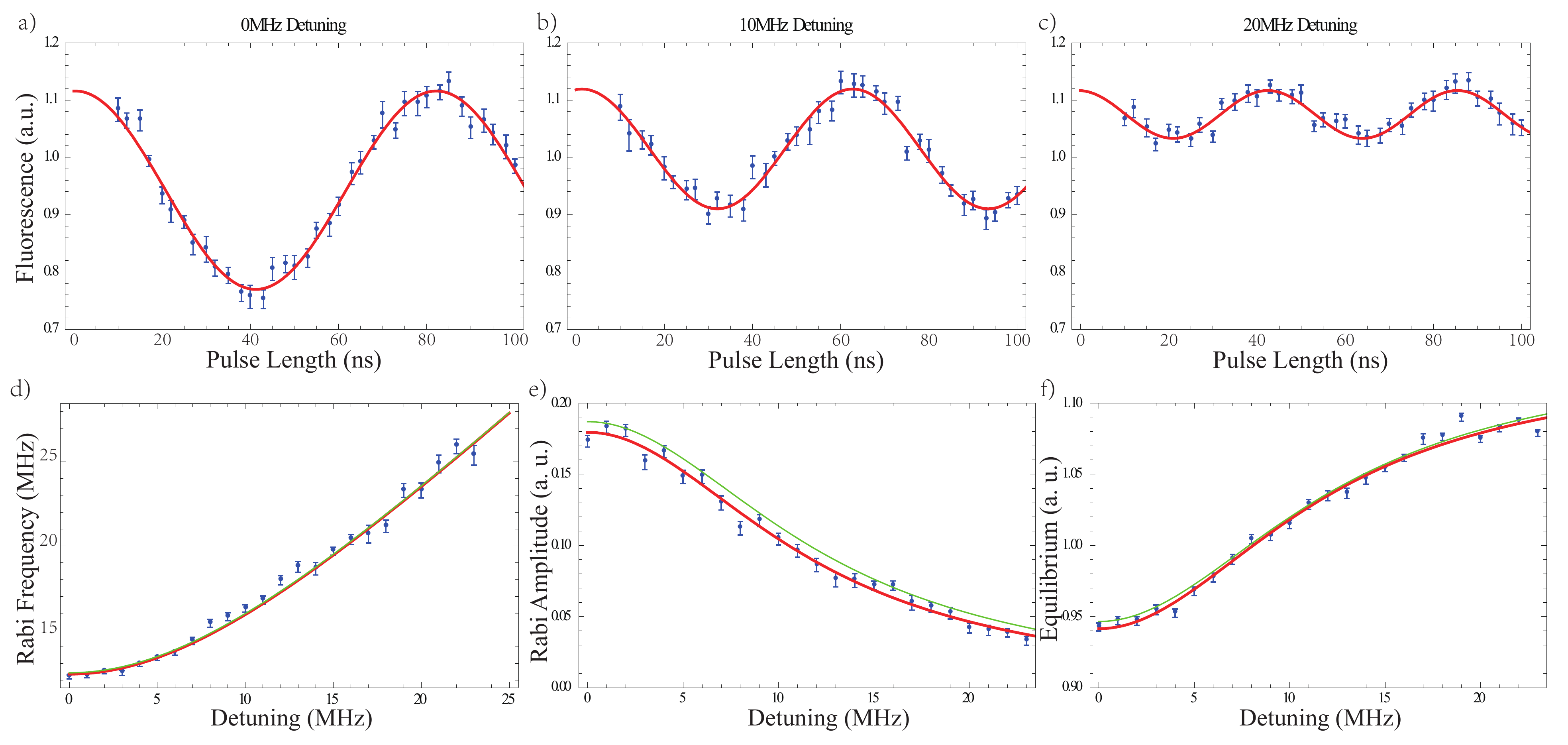}
	\caption{\textbf{a)}, \textbf{b)} and \textbf{c)} Rabi oscillation at 0 MHz, 10 MHz, and 20 MHz detuning. \textbf{d)}, \textbf{e)} and  \textbf{f)} Fitted frequencies, amplitudes and equilibrium positions at different detuning. Error bars come from standard error of fitted parameters. Red curve is a fit to the theory and green curve is the theoretical prediction based on our on-resonance Rabi data.}
\end{figure}

\subsection{Single Channel Nonlinearity of IQ Modulator}
One usual systematic error in our system is single channel nonlinearity of the IQ modulator. It means the microwave output amplitude is not always proportional to voltage input in I/Q channel. A Rabi experiment with fixed pulse length and scanning microwave amplitude is used to calibrate this single channel nonlinearity.

In the experiment, the microwave frequency is on resonance. The microwave pulse length is fixed to be 100~ns and different DC voltages are sent to I channel of the modulator. If there is no single channel nonlinearity, the microwave output amplitude should be proportional to the DC voltage on I channel, and so also the Rabi frequency. A sinusoidal signal in fluorescence level is expected.

\begin{figure}[h!]
	\includegraphics[width=3.4in]{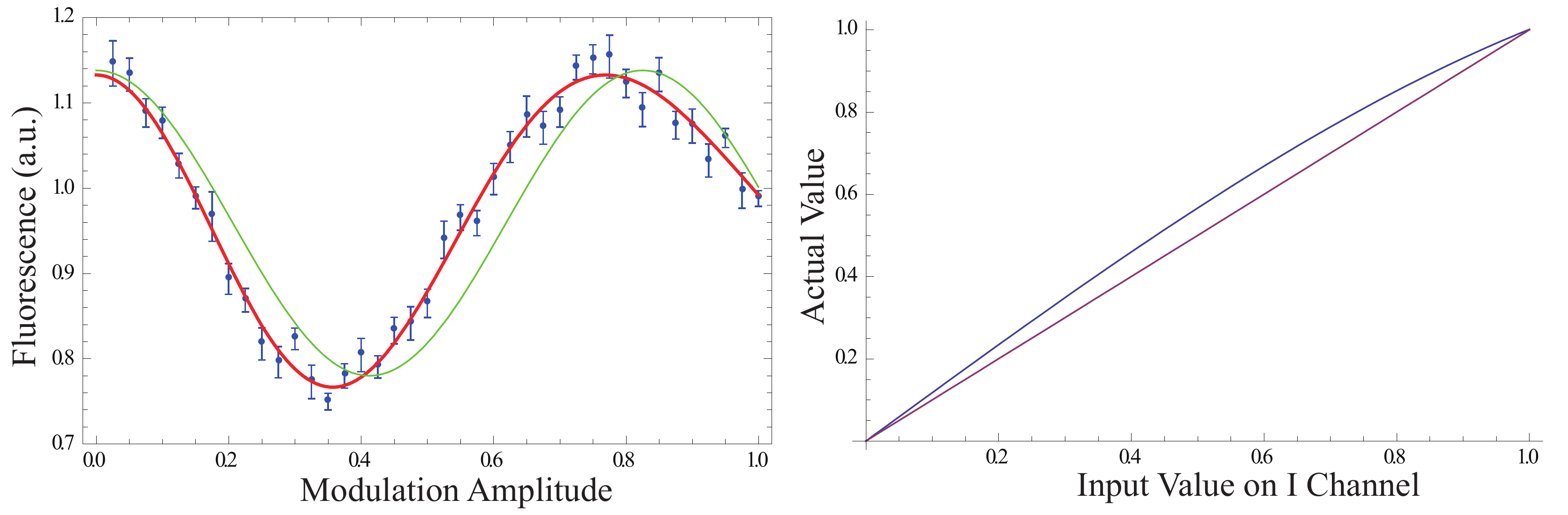}
	\caption{\textbf{a)}Rabi experiment with fixed microwave pulse length of 100ns and scanning DC voltage input to I channel. Green curve is expectation values without single channel nonlinearity, and red curve is the fit using 1st order nonlinearity.\textbf{b)}The calibrated actual value at different input value according to fitted parameters  $a=1.17, b=-0.17$. Clearly The actual value is larger than expectation, resulting in larger solid angle.}
	\label{figNonlinear}
\end{figure}

As shown in \figref{figNonlinear}, the green curve is the perfect sinusoidal function we expect, but the data points are clearly off by a certain amount. So the data is fitted to nonlinear function $a i+b i^3$ instead. The fit parameters are $a=1.17, b=-0.17$.

\subsection{Microwave Power Stability}

\begin{figure}[h!]
	\includegraphics[width=3.4in]{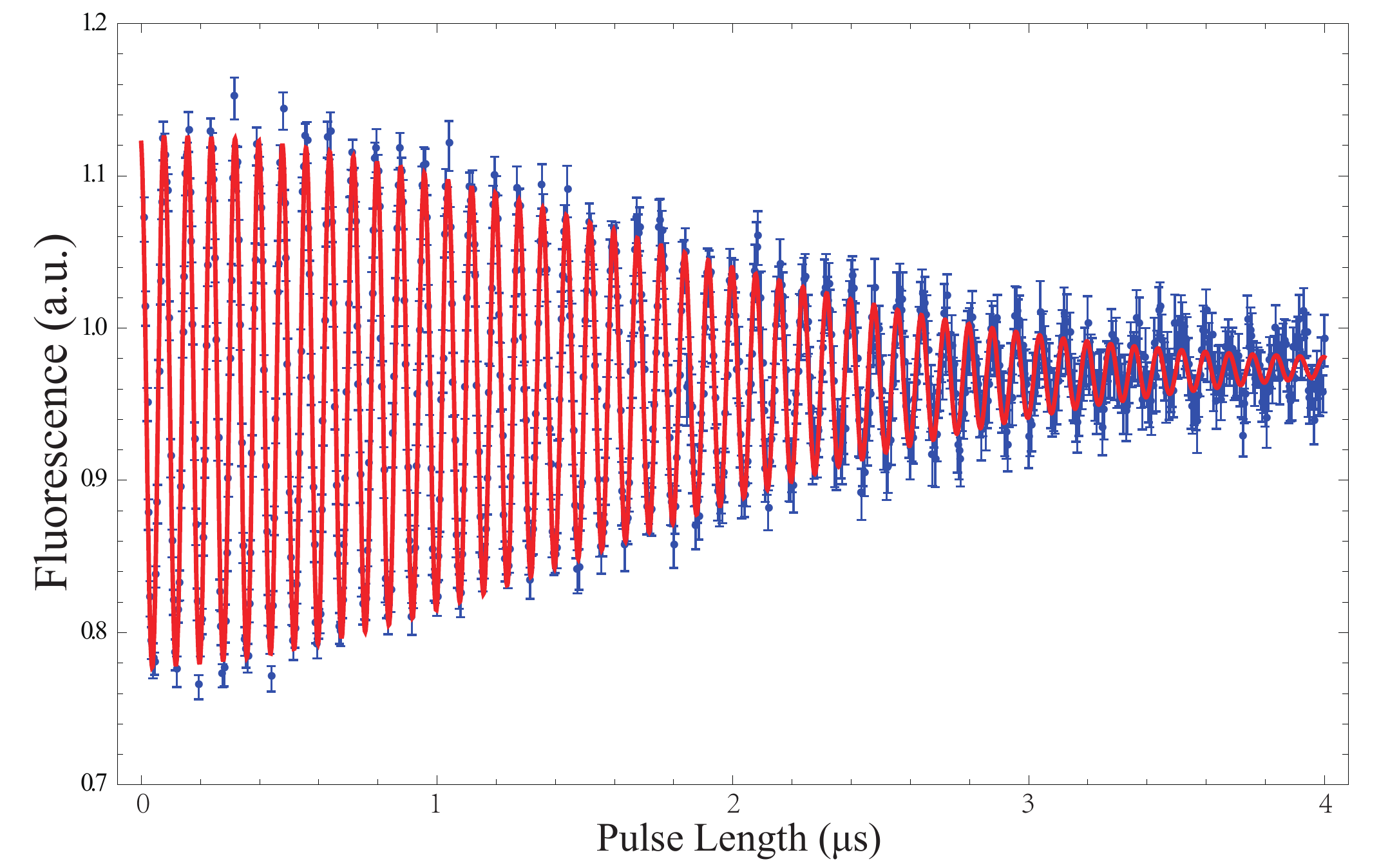}
	\caption{Rabi Oscillation up to long times. The fitted time scale of decay is $T_2^\rho = 2.2\mu s$.}
	\label{figLongR}
\end{figure}

A simple long time Rabi experiment was done to test the microwave power stability. If the microwave amplitude has a probability distribution around the expected value, the Rabi signal will decay as pulse length increases, which is shown in \figref{figLongR}.

The data is taken at the same microwave power as the Berry phase experiment (12.5MHz), and the total experiment time is comparable to Berry phase experiment. From the fit parameter we get the time scale of decay $T_1^\rho = 2.2 \mu s$, larger than the $T_2^*$ (about $0.8\mu s$ for this NV). This means that we do have fluctuation in microwave amplitude, but the overwhelming fluctuation is still the fluctuation in resonance frequency.

\section{Adiabaticity}

The Berry phase theory is based on adiabatic approximation. The Berry phase is not robust to transitions between $\ket{0}$ and $\ket{1}$ state, and there could be a discrepancy from the theory if the transition probability is not negligible.

\begin{figure}[h!]
	\includegraphics[width=3.4in]{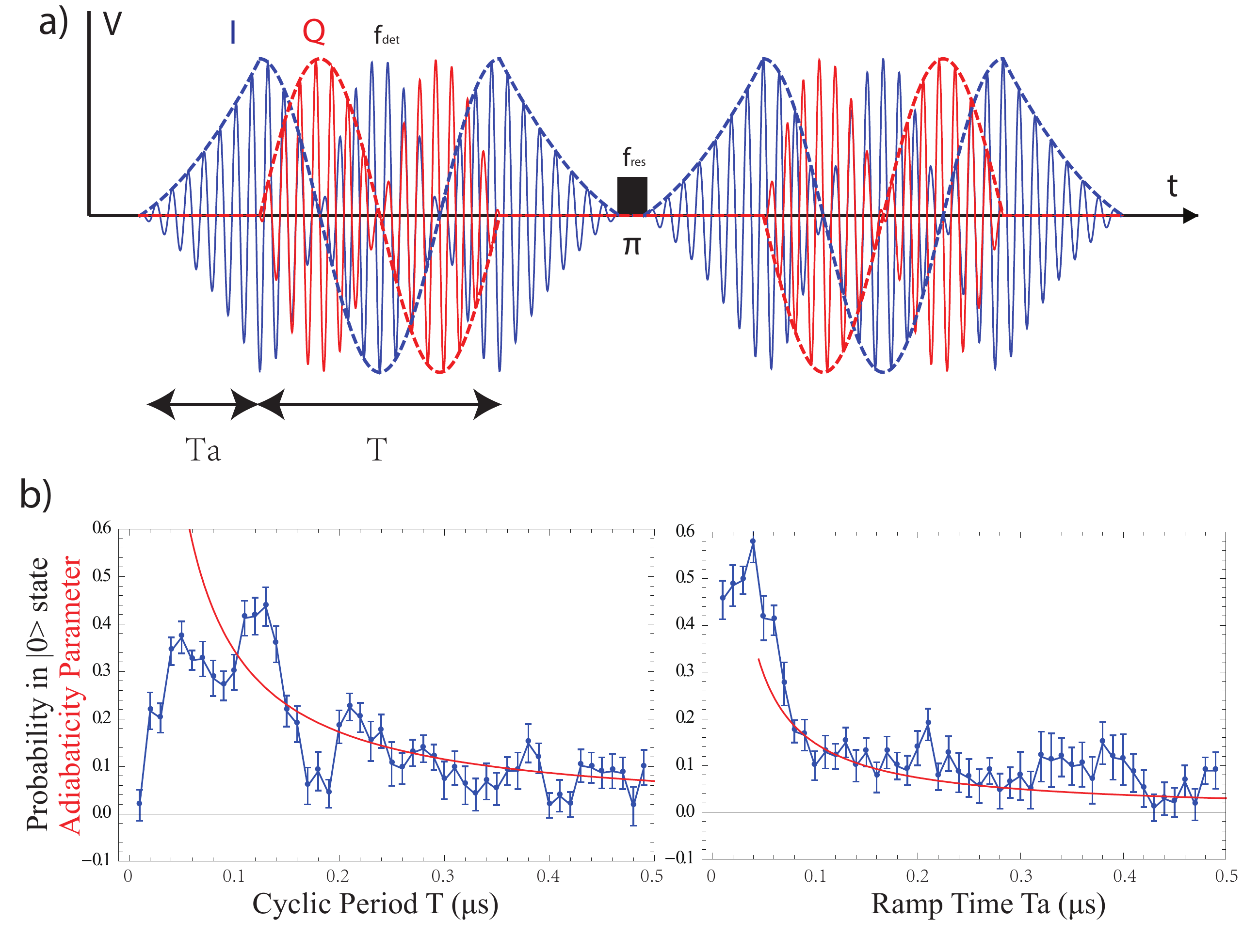}
	\caption{\textbf{a)} Schematic Microwave Sequence. \textbf{b)} Probability in $\ket{0}$ state vs period of cyclic motion T and  ramping time of microwave power $T_a$. Red curves are calculated adiabaticity parameter. Data was taken at 5 MHz detuning.}
	\label{figAdia}
\end{figure}

In order to check our adiabaticity, almost the same waveform as Berry phase experiment (N=2) was used. The only difference is that the first $\pi/2$ pulse and the last $\pi/2$ pulse were removed \figref{figAdia}. If the adiabatic condition is good, the spin should stay in $\ket{0}$ state before the $\pi$ pulse and stay in $\ket{1}$ state after the $\pi$ pulse. The probability in state $\ket{0}$ at the end of the sequence should be close to 0. This probability in state $\ket{0}$ for both scanning ramping up/down time Ta and the scanning cyclic period T is shown in \figref{figAdia}.

The data is taken at 5MHz detuning, 12.5MHz Rabi frequency. Adiabaticity parameter (shown as red curves) was calculated according to T($T_a$) using Equations:

\begin{align}
A & =\frac{\frac{2\pi}{T}sin\theta}{2\sqrt{\Delta^2+\Omega^2}}  & \text{(for cyclic motion)}\\
A & =\frac{1}{2\Delta T_a} \sqrt{\frac{\Omega^2}{\Delta^2+\Omega^2}} & \text{(for ramping)}
\end{align}

In our Berry phase experiments (10MHz detuning, 12.5MHz Rabi frequency for amplitude scan), we chose $T\ge800ns$, $T_a\ge350ns$. (See Table I for details.) We believe we were well within the adiabatic regime.

\begin{table}
\begin{tabular}{| c | c | c |}
  \hline
  Parameter & Experiment & Value \\
  \hline                       
  \multirow{3}{*}{Period of cycle $T$} & $\Omega$ scan N=2 & $1\mu s$ \\ \cline{2-3}
   & $\Omega$ scan N=3 & $0.8\mu s$ \\ \cline{2-3}
   & N scan & $0.8\mu s$ \\ 
  \hline 
  \multirow{3}{*}{Ramp time $T_a$} & $\Omega$ scan N=2 & $0.45\mu s$ \\ \cline{2-3}
     & $\Omega$ scan N=3 & $0.35\mu s$ \\ \cline{2-3}
     & N scan & $0.45\mu s$ \\
  \hline 
  \multirow{3}{*}{Spin Echo Sequence time $\tau$} & $\Omega$ scan N=2 & $4\mu s$ \\ \cline{2-3}
       & $\Omega$ scan N=3 & $4\mu s$ \\ \cline{2-3}
       & N scan & $20\mu s$ \\
  \hline  
\end{tabular}
\caption{Time parameters used in Berry phase experiments}
\end{table}


\section{Cancellation of Berry phase}

By flipping the direction of phase modulation at the 2nd half of the Berry phase sequence, the accumulated Berry phase survives while the dynamic phase gets cancelled by the spin echo. This implies that the Berry phase will be cancelled if the direction of phase modulation is not flipped. The cancellation of Berry phase is proved experimentally by applying the same sequence as amplitude scan (N=2), but without flipping the direction of phase modulation in the 2nd half. 

The data shows no oscillation at all, meaning a perfect cancellation of Berry phase by spin echo. And this is the raw data from which the $N=0$ phase data in the main paper is extracted.
\begin{figure}[h!]
	\includegraphics[width=3.4in]{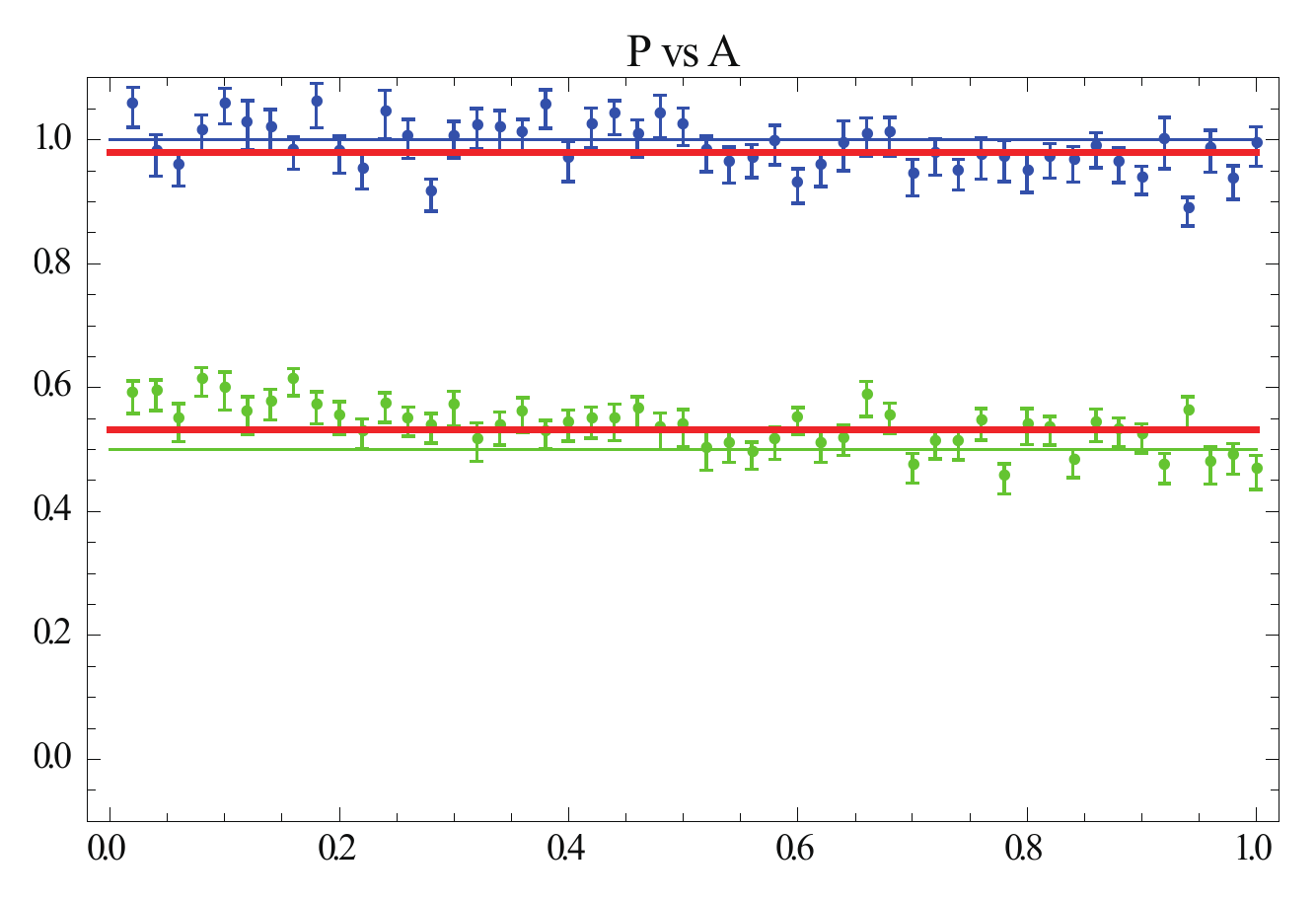}
	\caption{Cancellation of Berry Phase. We didn't flip the direction of cyclic motion of effective B field in the second half of spin echo. The Berry phase is cancelled just like the dynamic phase. }
	\label{figCancel}
\end{figure}

\section{Simulation of Berry Phase Experiment}

In the main text simulation of Berry phase experiment is shown in comparison to experimental data. The whole sequence is separated into small time steps ($\delta t << \frac{1}{\Omega}, \frac{1}{\Delta}$), and the time-dependent Hamiltonian in the rotating frame is approximately constant during each time step. The quantum state is evolving as $\ket{\Psi(t+\delta t)}=exp(-i \frac{\hat{H}(t)}{\bar{h}}\delta t) \ket{\Psi(t)}$. We can simulate our experiment with this method. Any imperfections in our microwave drive field can be easily taken into account by modifying the Hamiltonian $\hat{H}(t)$.

\[
 \hat{H}(t) = \Delta \hat{S_z} + \Omega \overbrace{(1.17\times I(t)-0.17\times I(t)^3)}^{\mathclap{\text{single channel nonlinearity}}}\hat{S_x}
\]
\[
 - \Omega \overbrace{(1.17\times\underbrace{(1.1Q(t))}_\text{amplitude imbalence}-0.17\times\underbrace{(1.1Q(t))^3}_{\mathclap{\text{amplitude imbalence}}})}^{\mathclap{\text{single channel nonlinearity}}}\hat{S_y}
\]

Parameters in single channel nonlinearity are extracted from experiment (see SI. III. C.), while parameter for amplitude imbalance is estimated from accuracy of our measurement of microwave amplitude. A microwave mixer is used to step down the frequency of microwave into the bandwidth of our oscilloscope, and the amplitude of microwave is measured with the oscilloscope.

The on-resonance microwave pulses for spin echo are assumed perfect, and decoherence is not taken into account because the sequence time $\tau<<T_2$.

Besides the simulation of Berry phase experiments (main paper and \figref{figDephasing}) and dynamic dephasing experiment (\figref{figDephasing}), we also simulate adiabaticity check (\figref{figAdia}) and cancellation experiment (\figref{figCancel}). They all agree with our corresponding experimental data.


\section{Estimating Berry Phase Decay from Measured $T_2$}
The measured Berry phase in our experiment for scanning $N$ is given by,
\beq
\beta(\tau/2) = \int \Theta_C dN = \int_{0}^{\tau/2} \biggl( \doe{\Theta_c}{\Delta} \biggr)\biggl\rvert_{\Delta} \frac{1}{T} f(t') \, dt'
\eeq
where $T$ is the cycle time for the contour, $f(t)$ is a Gaussian, stationary, i.i.d. random process with mean zero corresponding to fast frequency fluctuations in $\Delta$. We can set $\eta = \biggl( \doe{\Theta_c}{\Delta} \biggr)\biggl\rvert_{\Delta} \frac{1}{T}$ which is a constant factor that only depends on experimentally measured parameters. At the end of a spin-echo sequence, the phase coherence averaged over all realizations of the random process $f(t)$ is given by,
\beq
\langle e^{- 2 i \beta(\tau/2) + i \beta(\tau) } \rangle = \exp \biggl[-2 \langle\beta^2(\tau/2)\rangle - \frac{\langle \beta^2 (\tau) \rangle}{2} + 2 \langle \beta(\tau/2) \beta(\tau) \rangle) \biggr]
\eeq
where we have made use of the fact that $\langle e^{-i x} \rangle = \exp (- \langle x^2 \rangle/2)$ for a Gaussian random process $x$. If we now assume that the noise spectrum $S(\omega) = S_0$ upto some cut-off frequency $\omega_c = 1/ \tau_c$ where $\tau_c$ is the correlation time for the process, we can show that,
\beq
\langle e^{- 2 i \beta(\tau/2) + i \beta(\tau) } \rangle = \exp (- \frac{N^4}{N_2^4} )
\eeq
where $N_2 = \frac{1}{\sqrt{\eta}}\frac{T_2}{T} \sim 1250$ was calculated  from the experimental parameters $T_2 = \bigl(\frac{S_0}{6 \pi \tau_c^3} \bigr)^{(1/4)} \sim 200 \, \mu$s, $T = 0.8 \, \mu$s, $\Delta = 10$~MHz, $\Omega = 12.$~MHz, implying that the Berry phase decay predicted by the measured $T_2$ (fast frequency fluctuations in $\Delta$) is much slower than what we measure in the experiment.

\section{Simulation of Dynamic and Geometric Dephasing}

If the standard deviation of dynamic phase accumulates up to $\sqrt{2}$ radians, the amplitude of oscillation decays to 1/e. In our experiments, the spin echo sequence cancels most of the dynmic phase except the fluctuation of dynamic phase from one half to the other. The dynamic phase in one half is calculated by:
\begin{equation}
\beta_d = \sqrt{\Omega^2 + \Delta^2} \times\frac{t}{2}
\end{equation}
where $t= N^* T$, T is the period of cyclic motion. (See \figref{figDephasing}(a) right inset in comparison to Fig. 4(a) in main paper.)

The fluctuation of dynamic phase in one half is calculated as:
\begin{equation}
\sigma_{\beta_d} = \frac{\partial \beta_d}{\partial \Omega} \sigma_\Omega = \frac{\Omega^2}{\sqrt{\Omega^2 + \Delta^2}} \frac{N^* T}{2} (\frac{\sigma_\Omega}{\Omega}) = \sqrt{2} (rad)
\end{equation}
Here it is assumed that the fluctuation in $\Delta$ is negligible for such a short time scale. By plugging in microwave parameters and fitted $N^*$, the estimated relative error in $\Omega$, $\sigma_\Omega/\Omega$, is in $10^{-3}$ order of magnitude, which is in good agreement with our long time Rabi experiment (see Section III.D)

\begin{figure}[h!]
	\includegraphics[width=3.4in]{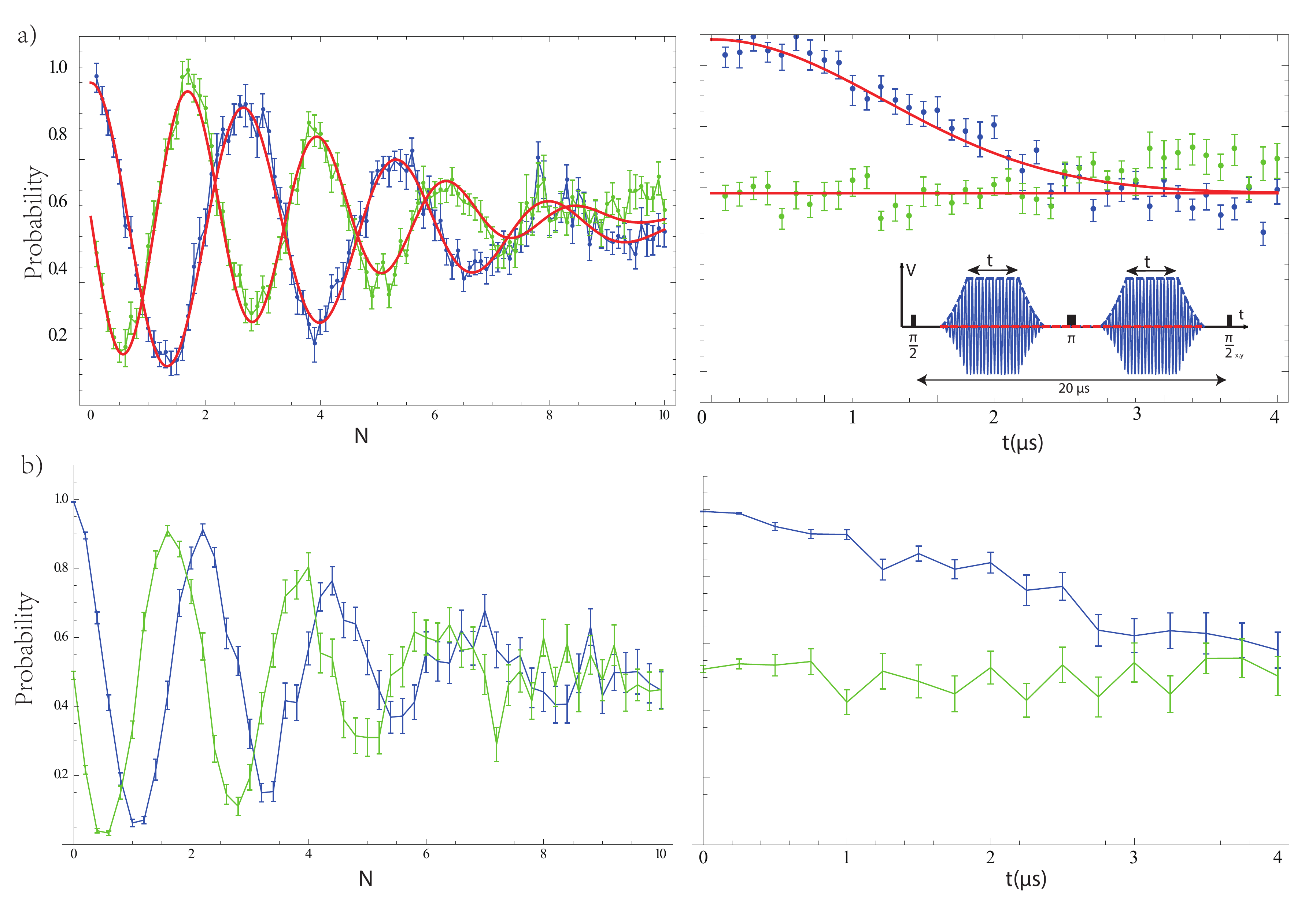}
	\caption{\textbf{a)} \textbf{(left)} N scan experiment (same as experiment of Fig. 4(c) in the main text) at $\Delta=10MHz$. Note that the x channel (blue) is oscillating slower than simulation, meaning a solid angle smaller than expected. This may be due to severe temperature drift on the day x channel data was taken. \textbf{(right)} Experimental verification of dynamic dephasing. The experimental sequence is shown in the inset. Here no geometric phase is introduced and thus the two halves of spin echo is designed to be identical. The quick decay of the data can only be explained by fluctuation of dynamic phase, which is most likely due to fluctuation of microwave amplitude. This decay shows comparable timescale is the N scan data, as $2 \tau = N T$, where T is the period of cyclic motion, defined in SI section IV. \textbf{b)} Numerical simulation of corresponding experiments in \textbf{a)}. We allowed the magnitude of the Rabi frequency $\Omega$ to fluctuate as a Gaussian random variable in either half of the sequence at a level of $8 \times 10^{-3}$.}
	\label{figDephasing}
\end{figure}

\bibliographystyle{apsrev4-1}

\end{document}